# Ultra-bright multiplexed energy-time entangled photon generation from lithium niobate on insulator chip

Guang-Tai Xue,[1,*] Yun-Fei Niu,[1,*] Xiaoyue Liu,[2,*] Jia-Chen Duan,[1,*] Wenjun Chen,[2] Ying Pan,[2] Kunpeng Jia,[1] Xiaohan Wang,[1] Hua-Ying Liu,[1] Yong Zhang,[1] Ping Xu,[3,1] Gang Zhao,[1] Xinlun Cai,[2] Yan-Xiao Gong,[1,†] Xiaopeng Hu,[1,‡] Zhenda Xie,[1, §] and Shining Zhu[1]

[1]*National Laboratory of Solid State Microstructures, School of Physics, School of Electronic Science and Engineering, College of Engineering and Applied Sciences, and Collaborative Innovation Center of Advanced Microstructures, Nanjing University, Nanjing 210093, China*

[2]*State Key Laboratory of Optoelectronic Materials and Technologies and School of Physics and Engineering, Sun Yat-sen University, Guangzhou 510275, China*

[3]*Institute for Quantum Information and State Key Laboratory of High Performance Computing, College of Computing, National University of Defense Technology, Changsha, 410073, China*

Abstract

High-flux entangled photon source is the key resource for quantum optical study and application. Here it is realized in a lithium niobate on isolator (LNOI) chip, with $2.79\times10^{11}$ Hz/mW photon pair rate and $1.53\times10^{9}$ Hz/nm/mW spectral brightness. These data are boosted by over two orders of magnitude compared to existing technologies. A 130-nm broad bandwidth is engineered for 8-channel multiplexed energy-time entanglement. Harnessed by high-extinction frequency correlation and Franson interferences up to 99.17% visibility, such energy-time entanglement multiplexing further enhances high-flux data rate, and warrants broad applications in quantum information processing on a chip.

## I. INTRODUCTION

Lithium niobate (LiNbO$_3$, LN) crystal is known for its superior optical performance [1], such as low optical transmission losses, large electro-optical and second-order nonlinear coefficients, etc. Therefore, it has been used for the fabrication of top-notch optical devices for both classical and quantum information applications. The recent progress of thin-film LN on Insulator (LNOI) [2] technology enables revolutionary footprint reduction of the LN devices by over three orders of magnitude, and thus makes a magnificent step forward towards efficient on-chip photonic integration. Various high performance on-chip optical devices have been developed based on such LNOI chips, including low-loss waveguides [3-5], high-quality-factor micro-ring resonators [3, 5, 6], and high-speed electro-optic modulators [7, 8], for applications in second harmonic generation [9-11], optical frequency comb generation [12, 13], and supercontinuum generation [14, 15].

Actually, the conventional waveguide devices on bulk-crystal LN wafers, with much weaker mode-confinement compared to the LNOI devices, have already shown their high efficiencies and performances in the form of integrated quantum optical circuits [2, 16]. Quantum states can be generated with unparalleled brightness by spontaneous parametric down-conversion (SPDC), and further tailored and modulated by domain engineering [17-19] and electro-optic modulation [20-22]. Recently, an important progress has been reported, which is that the electro-optic modulators have been realized with over 100 GHz [7, 8] bandwidth on the LNOI chip. It is not only a breakthrough in the classical optical communication, but also offers ultimate single photon switching power that can be matched with on-chip photon flux in the order of ~100 GHz. From the nonlinear optical point of view, it is natural to expect, with the new LNOI technologies, to push the photon state generation and processing efficiency to that unprecedented level, fulfilling the requirements of the next-generation quantum optical integration. A high-brightness entangled photon source is one of these key requirements, because it is basis for high-data-rate qubit generation, communication, and processing.

Here we demonstrate the first multiplexed energy-time entangled photon generation from a domain-engineered LNOI chip, with an ultra-high photon pair generation rate of $2.79 \times 10^{11}$ Hz/mW that is compatible with current LNOI electro-optic modulators. Small group velocity dispersion (GVD) is engineered in the type-0 quasi-phase-matching (QPM) [23]

SPDC process in a broad bandwidth of 130 nm, enabling eight-channel wavelength multiplexing. Franson interference is performed at each wavelength channel with visibilities over 97 %, and the maximum visibility exceeds 99.17 %. Our result shows the first high-data-rate photon source on LNOI chip, with both high photon flux and multiplexed energy-time entanglement, and found the basis for the large-scale high-density quantum information encoding, which is key to increase the efficiency for quantum cryptography [24, 25] and frequency coding [26-30], to fulfill the need of large-scale quantum communication, quantum information processing, and quantum simulation.

## II. DESIGN AND FABRICATION OF THE DOMAIN-ENGINEERED LNOI CHIP

Photon bandwidth is the key resource for wavelength multiplexing. Here for the type-0 QPM SPDC source in our design, a large bandwidth can be achieved by a combination of dispersion and domain engineering, enabling QPM around the zero GVD point [31, 32] (see the APPENDIX for details). The boundary conditions of the electromagnetic field determine its effective wave vector and corresponding dispersion as well as the GVD [33, 34]. For the waveguide, its boundary conditions are mainly determined by the structure and size. In contrast to the conventional LN waveguides, the LNOI waveguides can have high refractive index contrast and sub-micron structures. The schematic of the LNOI waveguide cross-section and simulation of the mode profiles is given in Fig. 1(a). By utilizing the software of COMSOL, we numerically simulate the GVD of light transmitting in the waveguide in the wavelength range from 1110 nm to 1570 nm. The simulated GVD for this structure is plotted in Fig. 1(b) (red). We can see that the GVD is very close to zero in the wavelength range from 1200 nm to 1570 nm, which is far from the zero GVD point around 1.92 μm in bulk LN crystal as shown in Fig. 1(b) (blue). At the working wavelength of our source, namely, 1475 nm, the GVD is about -60 $fs^2$/mm.

As shown in Fig. 1(c), the poling structure is designed for degenerate type-0 SPDC process from 737.5 nm pump to 1475 nm biphotons with a poling period of 4 μm. The fabrication process of the domain-engineered LNOI chip is illustrated in Fig. 1(f). We start from the commercial ion-sliced X-cut LNOI wafer (NANOLN) with a 600-nm lithium niobate film on 2-μm silica with 0.50-mm silicon substrate. First, the nichrome comb electrodes and the contact Au pads shown in Fig. 1(d) are fabricated by electron beam lithography (EBL), electron beam evaporation and lift off process. Then we use electrical field poling technique [35] to fabricate the domain structures. And several high-voltage pulses are applied to perform periodic domain inversion process. A confocal microscopic is used to monitor the poling quality on the LNOI chip. With image shown in Fig. 1(e), a duty cycle of about 50% is achieved, showing a high-quality fabrication. Next, the electrode is removed and a second EBL process is performed to define the photoresist for waveguide pattern. Finally, the waveguide structure is fabricated over the periodically poled LNOI (PPLNOI) by Inductively Coupled Plasma (ICP) etching process which is used to etch away 350-nm-thick lithium niobate film. The chip length is ~8 mm, and the poling region length is 6 mm. The waveguide width is linearly tapered from 1.4 μm to 4.5 μm at the output port, and leads to a waveguide-to-fiber-coupling efficiency of about 25 % at 1550 nm.

## III. SOURCE CHARACTERIZATION

We first test the PPLNOI waveguide by second harmonic generation (SHG) to find the optimal QPM wavelengths. A tunable cw laser (Santec TSL-550) is used as the fundamental light with power set to 0.7 mW. By tuning its wavelength, a maximum SHG light power of 4 μW is obtained after two short-pass filters (Thorlabs FESH0950) and the corresponding fundamental light wavelength is 1471.52 nm. The normalized SHG conversion efficiency is calculated to be 2270 %/W/cm$^2$.

Based on the SHG test, we set the pump wavelength at 735.76 nm for a reverse process in SPDC to the wavelength degeneracy around 1471.52 nm. The experimental setup of the energy-time entangled photon pair source is shown in Fig. 2(a), and the 735.76 nm pump light is from a tunable cw Ti: sapphire laser (M2 Sols). The PPLNOI chip is located on temperature-controlled mount to keep the QPM temperature at 23.0 °C. We use free-space coupling for both the input and output sides of the PPLNOI chip. For the output side, the entangled photons are coupled into a single-mode fiber after passing through a long-pass filter (Thorlabs FELH0950).

We build a high-performance wavelength-division multiplexing (WDM) filtering system to multiplex the energy-time entanglement, with schematic shown in Fig. 2(b). The SPDC light is directed through a single-mode fiber to collimator A for free-space output. A set of half-wave plate (HWP) and quarter-wave plate (QWP) is used to optimize the polarization for high diffraction efficiency on the following reflective diffraction grating G1. The diffracted SPDC light is then directed through an focusing lens in a 2-$f$ imaging setup, so that different frequency components are spatially chirped in the focal plane. A D-shaped mirror is used to reflect back the low-frequency components, and they are separated from the high-frequency components at the frequency degeneracy in the SPDC spectrum. The high-frequency components are collimated by the second lens for single-mode fiber coupling to collimator B, via a tunable reflective mirror M placed with separation of $f$ to the lens. The low-frequency components are directed back with small angle to the input direction in the beam height, and further diffracted by a second diffraction grating G2, for single-mode fiber coupling to collimator C. With angular dispersions, both collimator B and C receive 0.8-nm narrowband wavelength bins of the SPDC, and such wavelength multiplexing can be continuously turned by changing the angle of M and G2, respectively. This WDM filtering system covers a broad spectral range from 1350 nm to 1600 nm. A 3-dB insertion loss is measured for each output, with a high extinction ratio exceeding 60 dB. Utilizing this tunable WDM filtering system, we are able to pick entangled photon in any wavelength-bin pairs for measurement.

We use a coincidence measurement setup for the correlation measurements, as shown in Fig. 2(c). The single-mode-fibers coupled photon pairs are directed to and detected by a pair of superconducting nanowire single photon detectors (SNSPDs). Fiber polarization controller are used to optimize the polarization for highest detection efficiency of the SNSPDs. The overall efficiency of the SNSPDs is around 70 % including the fiber coupling, with dark counts of 3500 Hz. Coincidence measurement is performed by the time-correlated single-photon counting (TCSPC) module (Picoquant PicoHarp 300).

We first measure the SPDC generation rate by directly connecting fiber port O in Fig. 2(a) to fiber ports $D_1$ and $D_2$ of the coincidence counting system through a 50:50 fiber beam splitter. When the pump power coupled into the waveguide is 27.8 nW, a coincidence counting rate of $R_{cc}$ = 4792 Hz is measured, with the accidental coincidence counting rate $R_{ac}$ = 8 Hz. The ratio of coincidence to accidental coincidence is as high as 599. The single photon rate at the two detectors are $R_1$ = 126 kHz, and $R_2$ = 295 kHz, respectively, with negligible dark count level of 3500 Hz. Thus the on-chip photon pair generation rate $N = 2.79 \times 10^{11}$ Hz/mW can be calculated from $N = R_1 \times R_2 / (R_{cc} - R_{ac})$. The heralding efficiency is calculated to be 3.8%, which is consistent with the coupling efficiency including all losses [36]. The heralding efficiency of our source is relatively low because of the low waveguide-to-fiber coupling efficiency and the transmission loss of ~ 10 dB in the fiber connection from the source to detectors. The full bandwidth generation rate is the highest to our knowledge among all photon pair sources reported up to date.

To measure the spectrum brightness of the source, we connect port O in Fig. 2(a) to port A of WDM filter in Fig. 2(b) and perform the measurement around the frequency degeneracy. When the pump power coupled into the waveguide is 21.4 μW, the coincidence, accidental coincidence, and single counts are measured to be 1317 Hz, 12 Hz, 159 kHz, and 215 kHz, respectively, within the 0.8 nm bandwidth of WDM filter. A spectrum brightness of $1.53 \times 10^9$ Hz/nm/mW can be calculated, which is 1~2 orders of magnitude higher than the best result of the other photon pair sources (after normalization of the polling length) [20, 37- 41]. The above results fully reveal the advantage of the PPLNOI-based SPDC source with tight mode confinement. Actually, the waveguide cross-sectional area is 2 orders of magnitude smaller than that of the conventional lithium niobate waveguide, which is inversely proportional to the source efficiency (see the APPENDIX for details).

The SPDC spectrum is measured by heralded single photon counting, where port O is connected to a 50:50 fiber coupler, with one output used for the heralding signal and the other passing the WDM filter for the spectrum measurement. The D-shaped mirror is removed in this case, so that we can access the whole spectrum at port B. With a sweeping step of 10 nm, the SPDC spectrum shows a 130-nm full width at half maximum in Fig. 3(a), and the curve is fitted well with a sinc square function in good agreement with our simulation (as shown in the APPENDIX).

With the D-shape mirror inserted, we further measure the frequency correlation by WDM filter sweeping for both photons. A 2-dimensional sweep is performed with steps of 10 nm for B and C ports, respectively. The coincidence counting rate are plotted as a function of WDM filter wavelengths of ports B and C in Fig. 3(b). High-quality frequency anti-correlation is achieved, with a high rejection ratio of 32 dB for adjacent channels, and 40 dB for non-adjacent channels.

## IV. MEASUREMENT OF MULTIPLEXED ENERGY-TIME ENTANGLEMENT VIA FRANSON INTERFERENCE

The broad bandwidth and high-quality frequency correlation of the source enables the generation of wavelength-multiplexed energy-time entangled photon pairs, which can be harnessed using Franson interference [42] over each wavelength channel. We use only one unbalanced Michelson interferometer for the Franson measurement, where the multiplexed photon pairs are input from ports B and C of the WDM filter in Fig. 2(b) into ports $F_1$ and $F_2$ in Fig. 2(d) through fibers, respectively. Then they are polarization-controlled to be in orthogonal polarizations for the input of the Michelson interferometer, so that their relative phase difference can be easily stabilized. This Michelson interferometer is all-fiber based, and is form by a 50:50 fiber coupler, with a pair of fiber Faraday mirrors used to compensate any unwanted polarization change in the fiber for both long and short arms. The time imbalance of two arms $\Delta T = 1.5$ ns satisfies $T_{c1} \ll \Delta T \ll T_{c2}$, where $T_{c1}$ is the single-photon coherence time determined by the filtering bandwidth of 0.8 nm, and $T_{c2}$ is the coherence time of the pump laser. The interferometer is placed in a temperature-controlled box with the temperature control accuracy of 0.1 °C, and such accuracy is sufficient to change the relative phase delay $\phi$ between the two polarizations, which is distinguished by the stress-induced birefringence. Finally, the interferometer output is split and linked to ports $D_1$ and $D_2$ via a commercial coarse wavelength division multiplexer (CWDM) (with transition edge at 1471 nm) or a 50:50 fiber coupler for frequency nondegenerate or degenerate photon pairs, respectively. The coincidence window $\Delta t$ is set to 256 ps that is smaller than the time imbalance $\Delta T$, and hence coincidence events for both photons taking the long arm (L-L) and short arm (S-S) can be distinguished from those for one photon in long and the other in short (L-S) arms. Due to the energy-time entanglement, the events of L-L and S-S are indistinguishable, leading to an interference effect, namely, the Franson interference. Since the two photons pass the same interferometer, their path length difference automatically falls within the single-photon coherence length for maximum Franson interference visibility. The coincidence counting rate follows a $(1 + \cos\phi)$ variation [42], with $\phi$ depending on the phase difference of the L-L and S-S terms controlled by the interferometer box temperature.

The Franson interference are measured for each pair of the eight wavelength channels, as marked $\alpha_j, \beta_j, (j = 1, 2, \cdots, 8)$, with results shown in Fig.4 (b)-(i), respectively. All the curves are fitted well using sinusoidal functions, with the corresponding visibilities listed on top which all exceed 97%. Although we only pick eight WDM channels for energy-time entanglement measurement, the energy-time entanglement can be actually multiplexed in much denser spacing, and with the 0.8-nm WDM capability, over 100-channel pairs can be sliced within the 130-nm SPDC bandwidth. Together with frequency correlation measurement, these results are sufficient to demonstrate the capability LNOI-based photon source for high-flux wavelength-multiplexed energy-time entanglement generation in high fidelity, in the view of quantum magic bullet [43].

## V. DISCUSSION AND CONCLUSION

In summary, we demonstrated the first multiplexed energy-time entanglement generation in a domain-engineered LNOI chip. Ultra-high generation rate is measured up to $2.79\times10^{11}$ Hz/mW, which is the highest to our knowledge in all photon pair sources reported so far. Such high photon rate matches the potential photon switching capability on the order of 100 GHz, which is also possible on the same LNOI chip, for source-integrated quantum information processing. Eight-channel wavelength multiplexing is achieved with a spectral brightness of $\sim10^9$ Hz/nm/mW, and a frequency correlation of 32 dB extinction. Franson interference is observed for all the wavelength channels, and the interference visibilities are all measured to be over 97 %, showing high-quality energy-time entanglement. Here we focus on the entangled photon source performance from the LNOI chip, and use off-chip elements for the state processing and measurement. However, all the devices are ready for the on-chip processing, including on-chip WDMs and Franson interferometers. Hence, our approach is more than a stand-alone entangled photon source, but a breakthrough towards full-function integration of a quantum optical circuit. With small footprints of all the devices including such high-performance photon source, high integration density can be expected with superior photon flux and thus data rate in LNOI, for the next generation of quantum information processing chip, including quantum communication, computation, and simulation applications.

After we submitted this work, we became aware of a related publication of Ref. [44].

## ACKNOWLEDGMENTS

This work was supported by the National Key R&D Program of China (2019YFA0705000, 2017YFA0303700), Key R&D Program of Guangdong Province (2018B030329001), Leading-edge technology Program of Jiangsu Natural Science Foundation (BK20192001), National Natural Science Foundation of China (51890861, 11690031, 11621091, 11627810, 11674169, 91950206, 11974178), the Fundamental Research Funds for the Central Universities (021314380177).

*These authors contributed equally to this work

†[gongyanxiao@nju.edu.cn](mailto:gongyanxiao@nju.edu.cn)

‡[xphu@nju.edu.cn](mailto:xphu@nju.edu.cn)

§[xiezhenda@nju.edu.cn](mailto:xiezhenda@nju.edu.cn)

# APPENDIX: CALCULATION OF THE SPECTRUM AND GENERATIOIN RATE OF THE SOURCE

We consider a cw pump laser to travel along the waveguide with a length of $L$ in the $x$ direction. Since the waveguide is fabricated to be single mode for the signal and idler fields, and the pump (p) coupling is also optimized to excite the fundamental mode in the waveguide, the interaction Hamiltonian for the type-0 QPM SPDC can be treated in a single mode and written as [45]

$$H_I(t) = \varepsilon_0 \int_L dx \chi^{(2)}(x) \hat{E}_p^{(+)}(x,t) \hat{E}_s^{(-)}(x,t) \hat{E}_i^{(-)}(x,t) + \text{H.c.}, \tag{A1}$$

where H.c. denotes the Hermitian conjugate part. The negative parts of the signal (s) and idler (i) field operators are given by

$$\hat{E}_j^{(-)}(x,t) = \int d\omega_j E_j^* e^{-i(k_j x - \omega_j t)} \hat{a}_j^\dagger(\omega_j), \quad j = s, i, \tag{A2}$$

where $E_j = i\sqrt{\hbar \omega_j / [4\pi \varepsilon_0 c n(\omega_j)]}$. The positive part of the pump (p) field operator is written as the classical field amplitude

$$E_p^{(+)}(x,t) = E_p e^{i(k_p x - \omega_p t)}. \tag{A3}$$

Due to the periodic poling, the second-order nonlinear susceptibility can be written in the following Fourier series

$$\chi^{(2)}(x) = d \sum_m f_m e^{-i G_m x}. \tag{A4}$$

where $d = d_{33}$ is the effective nonlinear coefficient, and $G_m = 2m\pi/\Lambda$ is the $m$th-order reciprocal with $\Lambda$ denoting the poling period and the corresponding Fourier coefficient is written as $f_m = 2/(m\pi)$ for a duty cycle of 0.5. Here we use the first order reciprocal, and thus we only keep the term of $m = 1$ and thus can write the QPM condition as

$$\Delta k = k_p - k_s - k_i - \frac{2\pi}{\Lambda} = 0. \tag{A5}$$

With analogy calculations made in Ref. [45], we can write the frequency-degenerate SPDC state up to first-order perturbation term as

$$|\Psi\rangle = |\text{vac}\rangle + A \int d\nu\, h(L\Delta k) \hat{a}_s^\dagger(\omega_{s0} + \nu) \hat{a}_i^\dagger(\omega_{i0} - \nu) |\text{vac}\rangle, \tag{A6}$$

where we define a frequency detuning $\nu$ from the central frequency and here we consider the frequency-degenerate case, namely, $\omega_{s0} = \omega_{i0} = \omega_p/2$, with $\omega_p$ denoting the pump frequency. The joint spectrum amplitude $h(L\Delta k)$ is determined by the phase matching function, written as

$$h(L\Delta k) = \exp\left(-i\frac{L\Delta k}{2}\right)\operatorname{sinc}\left(\frac{L\Delta k}{2}\right). \tag{A7}$$

The coefficient $A$ absorbs all the constants and slowly varying functions of frequency, given by

$$A = i\frac{d_{33}\alpha L E_p \omega_p}{2\pi c n_0}, \tag{A8}$$

where $n_0$ represents the refractive index of light at frequency $\omega_p/2$ and $\alpha$ is the overlap factor [46].

To further investigate the source spectrum, we expand signal and idler wave vectors around the central frequencies up to the second order in the detuning $\nu$, respectively,

$$k_s(\omega_s = \omega_p/2 + \nu) = \frac{n(\omega_p/2+\nu)}{c} \approx \frac{n_0\omega_p}{2c} + \frac{1}{u_0}\nu + \frac{\text{GVD}_0}{2}\nu^2, \tag{A9}$$

$$k_i(\omega_i = \omega_p/2 - \nu) = \frac{n(\omega_p/2-\nu)}{c} \approx \frac{n_0\omega_p}{2c} - \frac{1}{u_0}\nu + \frac{\text{GVD}_0}{2}\nu^2, \tag{A10}$$

where $u_0$ and $\text{GVD}_0$ are the group velocity and group velocity dispersion of light at frequency $\omega_p/2$, expressed as

$$u_0 = \left.\frac{d\omega}{dk}\right|_{\omega=\omega_p/2}, \tag{A11}$$

$$\text{GVD}_0 = \frac{d^2k}{d\omega^2} = \left.\frac{d}{dk}\left(\frac{1}{u}\right)\right|_{\omega=\omega_p/2}. \tag{A12}$$

Since the QPM condition given by Eq. (A5) is satisfied for signal and idler photons at $\omega_p/2$, we can obtain

$$\Delta k = k_p - k_s - k_i - \frac{2\pi}{\Lambda} \approx \text{GVD}_0\, \nu^2. \tag{A13}$$

Hence the spectrum amplitude given by Eq. (A7) can be estimated up to the second order in $\nu$ as

$$h(L\Delta k) = \exp\left(-i\frac{L\,\text{GVD}_0}{2}\nu^2\right)\operatorname{sinc}\left(\frac{L\,\text{GVD}_0}{2}\nu^2\right). \tag{A14}$$

With one photon triggered without spectrum analysis, the spectrum amplitude of the other photon can be obtained as

$$P(\nu) = |h(L\Delta k)|^2 = \operatorname{sinc}^2\left(\frac{L\,\text{GVD}_0}{2}\nu^2\right), \tag{A15}$$

and alternatively, as a function of wavelength it is written as,

$$P(\lambda) = \text{sinc}^2\left[2\pi^2 c^2 L\, \text{GVD}_0 \left(\frac{1}{\lambda} - \frac{1}{2\lambda_p}\right)^2\right], \tag{A16}$$

which is proportional to the coincidence count for spectrum measurement shown in Fig. 3(a). By utilizing the numerical simulation made in Fig. 1(d), and $\lambda_p = 735.76$ nm, $L = 6$ mm, $\text{GVD}_0 = -60$ fs$^2$/mm, we plot the spectrum amplitude against wavelength in Fig. 5, from which we can calculate the FWHM to be ~200 nm, which is larger than the experimental result of 130 nm due to the fabrication deviation with the theoretical design.

The generation rate can be obtained by calculating the trace of the two-photon term of the state given by Eq. (A6), having

$$R = |A|^2 \int d\nu\, |h(L\Delta k)|^2. \tag{A17}$$

Substituting Eqs. (A8), (A14), and $|E_p|^2 = 2P/(\varepsilon_0 n_p c S)$, we obtain

$$R = \frac{8 P d_{33}^2 \alpha^2}{3\varepsilon_0 c S \lambda_p^2 n_0^2 n_p} \sqrt{\frac{2\pi L^3}{|\text{GVD}_0|}}. \tag{A18}$$

where $S$ is the cross-sectional area of the pump and here it is the cross-sectional area of LNOI waveguide. We can see that the generation rate is inversely proportional to the area. Since the area of the LNOI waveguide is 1~2 orders of magnitude smaller than that of the conventional lithium niobate waveguide, the generation rate can be also increased 1~2 orders of magnitude. With $\lambda_p = 735.76$ nm, $P = 1$ mW, $S = 1.26$ μm$^2$, $\alpha^2 = 0.928$, $L = 6$ mm, and $d_{33} = 33$ pm/V, $GVD_0 = -60$ fs$^2$/mm, we obtain the photon pair generation rate as $R = 8.44 \times 10^{11}$ Hz/mW.

FIGURES

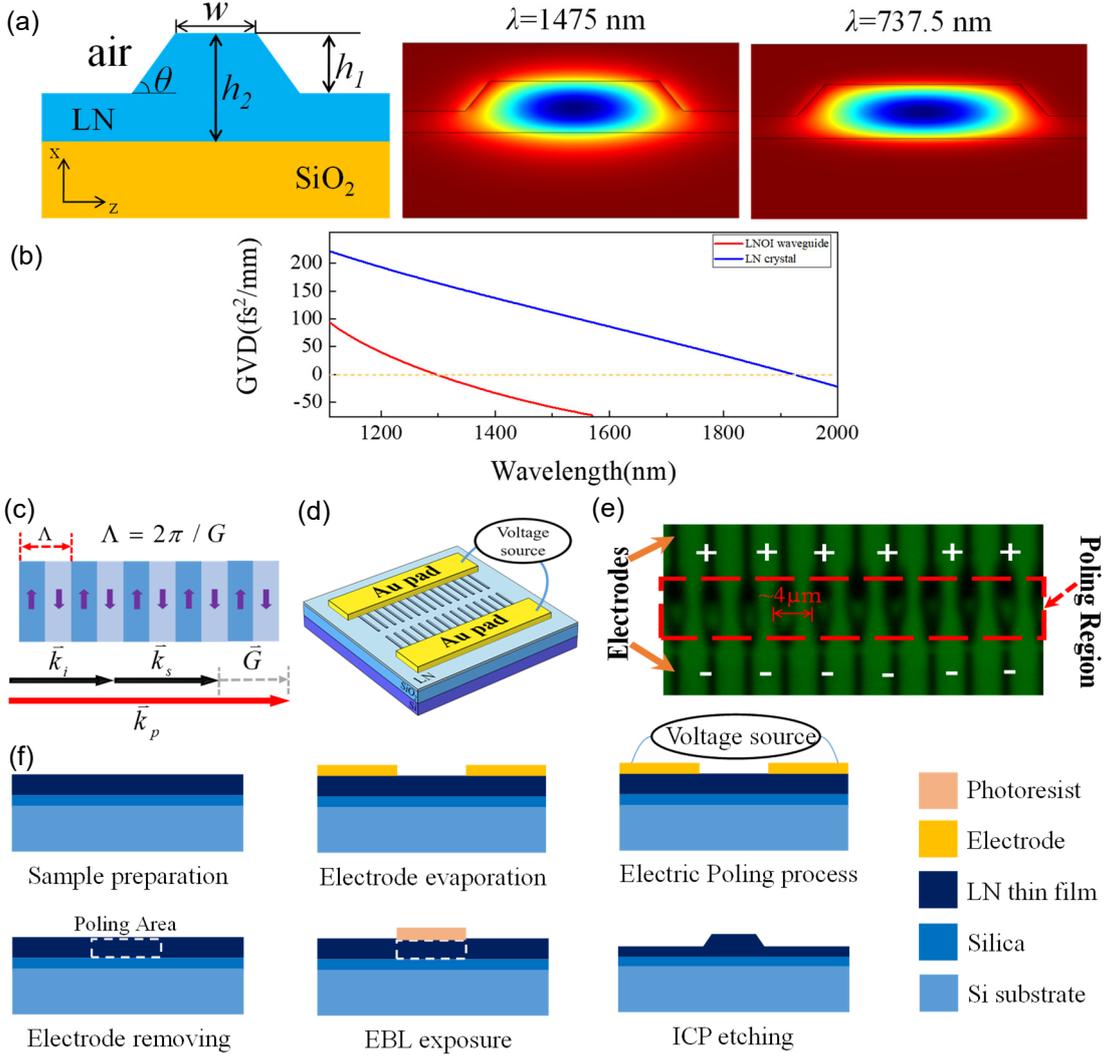

FIG. 1. Schematic of chip design and fabrication. (a) Cross-section of the waveguide and mode field simulation, with parameters $w$ = 1.4 μm, $h_1$ = 350 nm, $h_2$ = 600 nm and $\theta$ = 60°. (b) Simulation of the GVD for the LNOI waveguide (red line) and bulk lithium niobate crystal (blue line). (c) Quasi-phase matching geometry, with $p$, $s$, and $i$ denoting the pump, signal, and idler fields, respectively. (d) Electrode structure and electrical field poling process. High-voltage pulses are applied along the z-axis of the X-cut LNOI wafer. (e) Confocal laser scanning microscopy picture of domain structure. The black areas are domain walls and electrodes in the poling region and outside, respectively. (f) Schematic of the chip fabrication process.

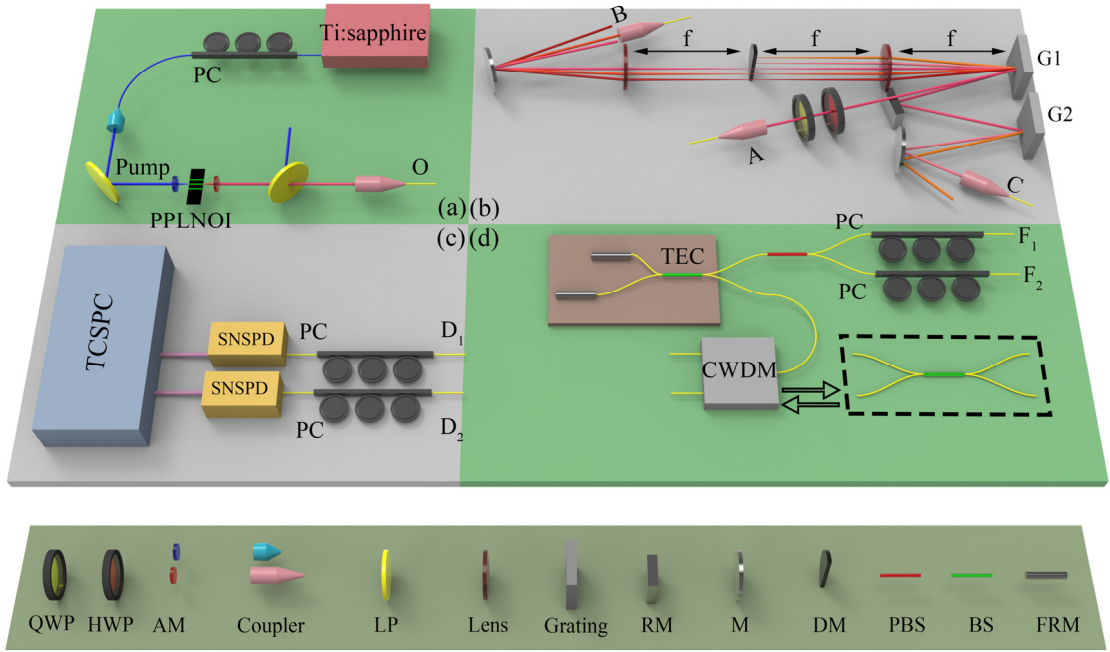

FIG. 2. Schematic of the experimental setup, including four parts: (a) the energy-time entangled photon pair source based on PPLNOI chip, (b) the tunable wavelength division multiplexing filtering system, (c) the coincidence count system, and (d) the Franson interferometer for testing energy-time entanglement. QWP: quarter wave plate; HWP: half wave plate; AM: aspheric lens; LP: long-pass filter; M: mirror; RM: rectangular mirror; DM: D-shaped mirror; PBS: polarizing beam splitter; BS: 50:50 fiber beam splitter; FRM: Faraday rotator mirror; PC: polarization controller; TEC: thermo-electric cooler; CWDM: a commercial coarse wavelength division multiplexer; SNSPD: superconducting nanowire single photon detector; TCSPC: time-correlated single-photon counting. The letters O, A, B, C, $D_1$, $D_2$, $F_1$ and $F_2$ represent the input or output ports of the four parts, which are separated and can be connected in different ways as needed.

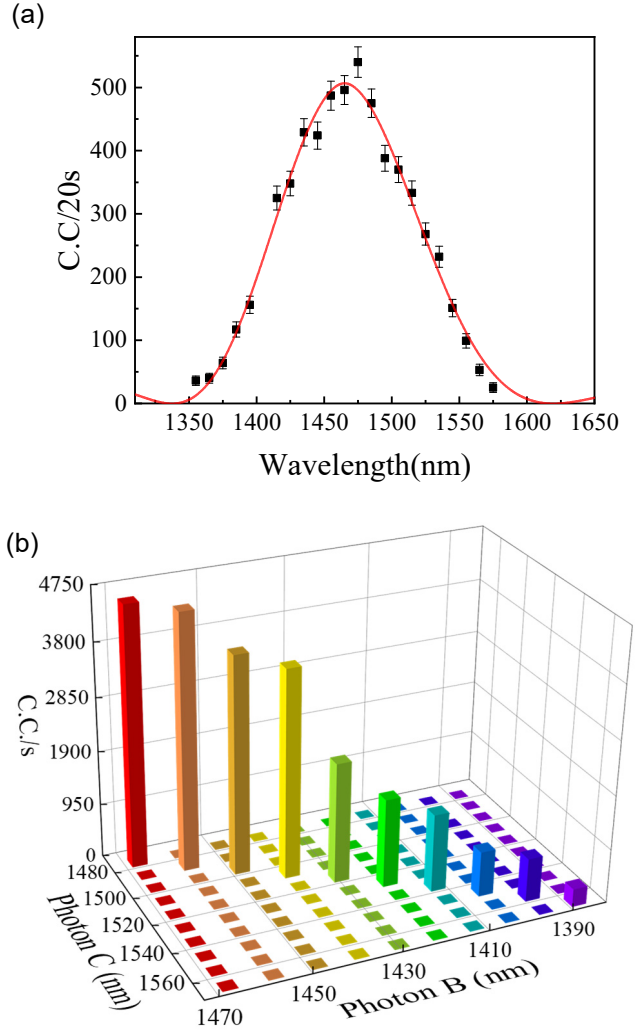

FIG. 3. Spectrum characterization of the source. (a) SPDC spectrum. Black points represent coincidence counts as a function of the filtering wavelength of one photon, with the other photon unfiltered. The error bars represent the square root of the counts. The solid curve is fitted with a sinc square function. (b) Frequency correlation measurement. Coincidence counts are recorded against the wavelengths of the photons from couplers B and C in steps of 10 nm.

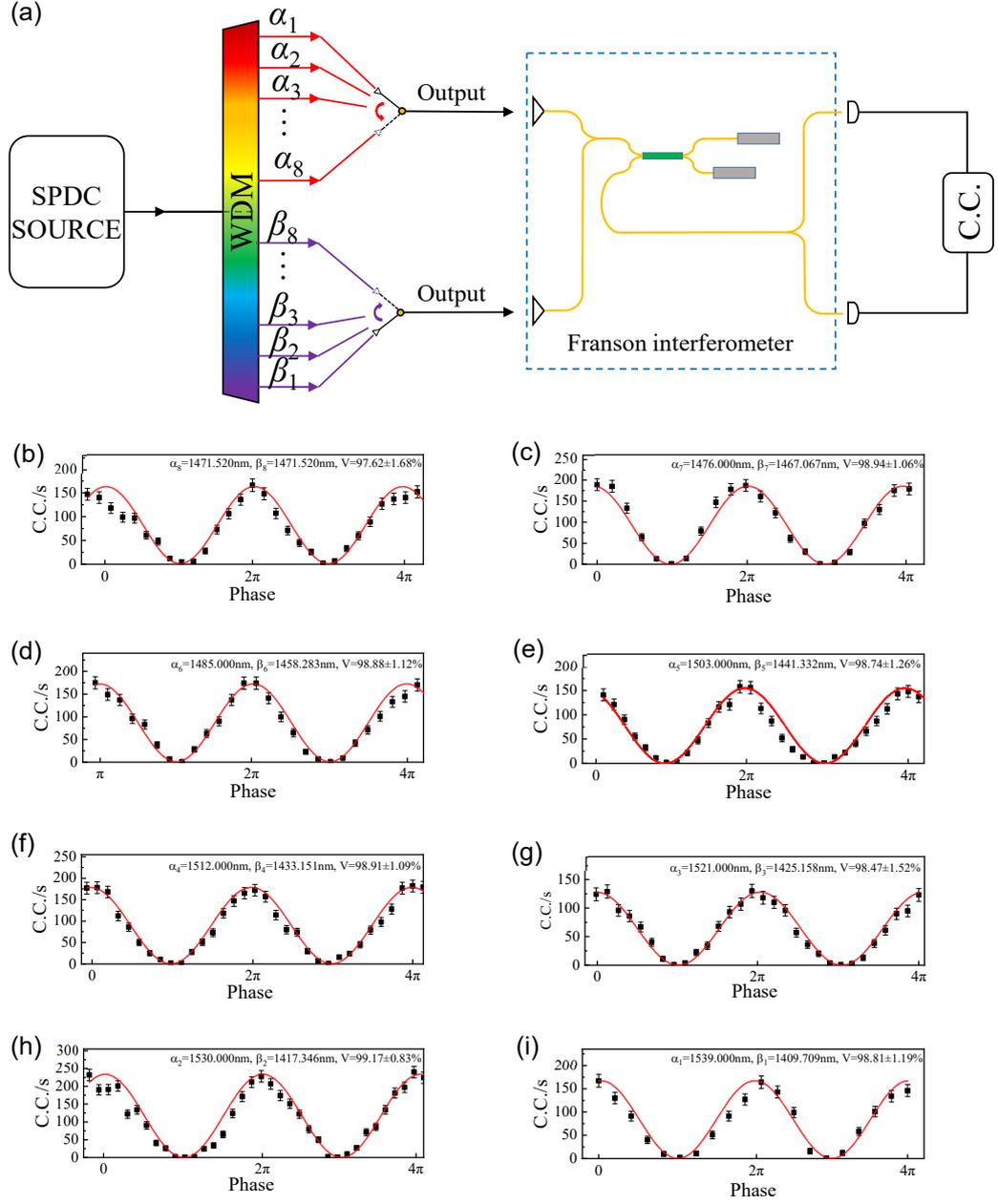

FIG. 4. Multiplexed energy-time entanglement characterization. (a) Illustration of the multiplexed energy-time entanglement generation from our source. Franson interferences are measured for each wavelength bin after the WDM filter. (b)-(i) Experimental results of the coincidence counts for entangled photons in eight pairs of wavelength channels $\alpha_j$, $\beta_j$, ($j = 1, 2, \cdots, 8$), as a function of the phase difference of both photons taking the long arm and both taking the short arm, with error bars denoting the square root of the counts. All the curves are fitted with sine-cosine functions, with the corresponding visibilities listed on top, where the errors are estimated assuming a Poisson fluctuation in the counts.

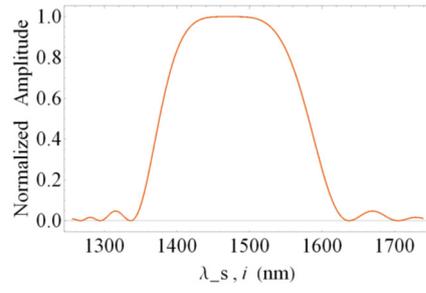

FIG. 5. Simulation of the source spectrum based on theoretical design.